\documentclass[aps,pra,twocolumn,showpacs]{revtex4}
\usepackage[dvips]{graphicx}
\usepackage{amsmath}
\usepackage{times}
\usepackage{psfrag}

\begin{document}

\title
{
First- and second-order superfluid--Mott-insulator phase transitions of spin-1 bosons
with coupled ground states in optical lattices
}

\author
{K.~V.~Krutitsky, M.~Timmer, and R.~Graham
}

\affiliation{
Fachbereich Physik der Universit\"at Duisburg-Essen, Campus Essen,
Universit\"atsstr. 5, 45117 Essen, Germany
}

\date{\today}

\begin{abstract}
We investigate the superfluid--Mott-insulator quantum phase transition
of spin-1 bosons in an optical lattice created by pairs of
counterpropagating linearly polarized laser beams,
driving an $F_g=1$ to $F_e=1$ internal atomic transition.
The whole parameter space of the resulting two-component Bose-Hubbard model is
studied. We find that the phase transition is not always second order
as in the case of spinless bosons,
but can be first order in certain regions of the parameter space.
The calculations are done in the mean-field approximation
by means of exact numerical diagonalization as well as within the framework of
perturbaton theory.
\end{abstract}


\maketitle

The superfluid--Mott-insulator quantum phase transition (SMQPT) of spinless bosons
in periodic lattices is a second-order transion, which is characterized by
a continuous variation of the order parameter $\psi$ from $\psi \ne 0$ (superfluid phase)
to $\psi=0$ (Mott-insulator phase) if the amplitude of the lattice potential
increases~\cite{Sachdev,Oosten}.
In our recent paper~\cite{KGPRA04}, we have shown that the SMQPT in a system of spin-1
bosons can be first order as well~\cite{FO}.
By means of numerical calculations within the framework
of the mean-field theory, it was found that in the case of $^{23}$Na the SMQPT
is second order if the number of atoms per lattice site $n=1,3$, and it is first order
for $n=2$. In the case of $^{87}$Rb, the SMQPT was found to be second order for $n=1,2,3$.
In the present work, we continue the study of Ref.~\cite{KGPRA04}. The main purpose
is to investigate the whole parameter space of spin-1 bosons and to find the regions
where the SMQPT is first and second order for arbitrary $n$.

We consider spin-1 neutral polarizable bosons,
possessing three Zeeman-degenerate internal ground and excited electronic states
characterized by the magnetic quantum number $m = 0,\pm 1$ ($F_g=F_e=1$)
in a $d$-dimensional ($d=1,2,3$) optical lattice. The lattice is assumed to be created
by $d$ pairs of counterpropagating linearly polarized laser waves running in $d$ orthogonal
directions and having different frequencies in different directions. The beams
propagating along the $3$-axis, which is chosen to be a quantization axis, are polarized
along the $1$- or $2$-axis, and the beams propagating along the $1$- and $2$-axis are
polarized along the $2$- and $1$-axis, respectively.

The running laser waves form left- and right-polarized standing waves with Rabi frequencies
$
\Omega_\nu(r_\nu)
=
\Omega_{0\nu}
\cos(k_L r_\nu)
$,
which couple internal ground and excited states by $V$- and $\Lambda$- transitions.
In order to avoid decoherence due to spontaneous emission,
the detunings $\Delta_i$ must satisfy the conditions $|\Delta_\nu| \gg \gamma$,
where $\gamma$ is the spontaneous emission rate.
If the laser intensities and the detunings are chosen in such a manner that
$
  \Omega_{0\nu}^2/\Delta_\nu
  =
  \Omega_{0}^2/\Delta
$,
$\nu=1,\dots,d$, the laser potential
acting on the atomic ground states is given by the matrix
\begin{equation}
\label{matrix}
V_L({\bf r})
=
\hbar
\frac{\Omega_0^2}{\Delta}
\sum_{\nu=1}^d
\cos^2(k_L r_\nu)
\left(
    \begin{array}{ccc}
       1 & 0 & 1 \\
       0 & 2 & 0\\
       1 & 0 & 1
    \end{array}
\right)
\;,
\end{equation}
which
determines the isotropic lattice potential with the period $\pi/k_L$ and at the same time
couples the atomic ground states with $m = \pm 1$. In the case of red detuning $\Delta<0$,
the system is described by the two-component Bose-Hubbard Hamiltonian~\cite{KGPRA04}
\begin{eqnarray}
\label{BHH}
&&
\hat H_{BH}
=
-
J
\sum_{<i,j>}
\left(
    \hat a_{0i}^\dagger
    \hat a_{0j}
    +
    \hat a_{\Lambda i}^\dagger
    \hat a_{\Lambda j}
\right)
\nonumber\\
&&
+
\sum_i
\left[
\frac{U_s}{2}
\hat n_i (\hat n_i-1)
+
U_a
\hat n_{0i} \hat n_{\Lambda i}
\right.
\\
&&
\left.
-
\frac{|U_a|}{2}
\left(
    \hat a_{0i}^\dagger
    \hat a_{0i}^\dagger
    \hat a_{\Lambda i}
    \hat a_{\Lambda i}
    +
    \hat a_{\Lambda i}^\dagger
    \hat a_{\Lambda i}^\dagger
    \hat a_{0i}
    \hat a_{0i}
\right)
-
\mu
\hat n_i
\right]
\,,
\nonumber
\end{eqnarray}
where $\mu$ is a chemical potential and the operator
$\hat n_i=\hat n_{0i}+\hat n_{\Lambda i}$.
The indices $i,j$ label the lattice sites, and
$\hat a_{\sigma i}$ is the Bose annihilation operator for the component $\sigma=0,\Lambda$
attached to the $i$th lattice site.
The tunneling matrix element
\begin{displaymath}
J
=
-\int
W_{i+1}({\bf r})
\left[
    -
    \frac{\hbar^2}{2M}
    \frac{\partial^2}{\partial {\bf r}^2}
    +
    V_{L0}({\bf r})
\right]
W_i({\bf r})
\,d{\bf r}
\;,
\end{displaymath}
where $W_i({\bf r})$ is the Wannier function for the lowest Bloch band of the potential
\begin{displaymath}
  V_{L0}({\bf r})
  =
  2\hbar\frac{\Omega_0^2}{\Delta}
  \sum_{\nu=1}^d
  \cos^2(k_L r_\nu)
\;,
\end{displaymath}
is a rapidly varying function of the laser intensity~\cite{Jaksch}.
The variations of the atomic interaction parameters
\begin{displaymath}
U_{s,a}
=
g_{s,a}
\int
W_i^4({\bf r})
\,d{\bf r}
\end{displaymath}
are much slower~\cite{Jaksch}.
The quantities $g_{s,a}$ describe the repulsive interaction of the condensate
atoms and the spin-changing collisions.
The parameter $U_s$ is positive, but $U_a$ can be either positive or negative depending
on the sign of $g_a$.

In order to investigate the SMQPT,
we employ the mean-field approximation~\cite{Sachdev,Oosten}
\begin{equation}
\label{mfa}
\hat a_{\sigma i}^\dagger
\hat a_{\sigma j}
\approx
\psi_\sigma
\left(
    \hat a_{\sigma j}
    +
    \hat a_{\sigma i}^\dagger
\right)
-
\psi_\sigma^2
\;,
\end{equation}
where $\psi_\sigma$ is the order parameter for Bose-Einstein condensation
in the component $\sigma=0,\Lambda$, which can be considered
as a real quantity.
In this approximation, the Bose-Hubbard Hamiltonian becomes local and every
lattice site is described by the Hamiltonian
\begin{eqnarray}
\label{HBH1}
\hat H_{BH}'
&=&
\hat H^{(0)}
+
\hat V
\;,
\\
\hat H^{(0)}
&=&
2 d J
\left(
    \psi_0^2
    +
    \psi_\Lambda^2
\right)
+
\frac{U_s}{2}
\hat n (\hat n - 1)
\nonumber\\
&&
+
U_a
\left[
    2
    \hat T_{1(2)}^2
    -
    \frac{\hat n}{2}
\right]
-
\mu
\hat n
\;,
\nonumber\\
\hat V
&=&
-
2 d J
\left[
    \left(
        \hat a_0^\dagger
        +
        \hat a_0
    \right)
    \psi_0
    +
    \left(
        \hat a_\Lambda^\dagger
        +
        \hat a_\Lambda
    \right)
    \psi_\Lambda
\right]
\;,
\nonumber
\end{eqnarray}
where the index $1(2)$ corresponds to $U_a<0$ ($U_a>0$).
In Eq.(\ref{HBH1}), we have omitted the site index $i$ and introduced the isospin
operator $\hat {\bf T}$ with the components
\begin{eqnarray}
\hat T_1
&=&
\left(
    \hat a_\Lambda^\dagger \hat a_0
    +
    \hat a_0^\dagger \hat a_\Lambda
\right)
/2
\;,
\nonumber\\
\hat T_2
&=&
i
\left(
    \hat a_\Lambda^\dagger \hat a_0
    -
    \hat a_0^\dagger \hat a_\Lambda
\right)
/2
\;,
\\
\hat T_3
&=&
\left(
    \hat a_0^\dagger \hat a_0
    -
    \hat a_\Lambda^\dagger \hat a_\Lambda
\right)
/2
\;,
\nonumber
\end{eqnarray}
which has the property
\begin{displaymath}
\hat {\bf T}^2
=
\frac{\hat n}{2}
\left(
    \frac{\hat n}{2} + 1
\right)
\;.
\end{displaymath}

If the tunneling $\hat V$ is negligible, the eigenstates of the Hamiltonian (\ref{HBH1})
are determined by $H^{(0)}$ and can be calculated analytically.
They are eigenstates $|n/2,{\cal M}\rangle$ of the isospin operator
with the corresponding eigenenergies given by
\begin{eqnarray}
\label{e}
E^{(0)}_{n/2,{\cal M}}
&=&
2 d J
\left(
    \psi_0^2
    +
    \psi_\Lambda^2
\right)
+
\frac{U_s}{2}
n (n - 1)
\nonumber\\
&&
+
U_a
\left(
    2
    {\cal M}^2
    -
    \frac{n}{2}
\right)
-
\mu n
\;,
\end{eqnarray}
where
$n$ is the number of atoms per lattice site and
${\cal M}=-n/2,\dots,n/2$ is the isospin projection on the direction $1(2)$
in the isospin space for the case of $U_a<0$ ($U_a>0$).
Note that $n/2$ plays the role of the isospin quantum number.
If $n$ is even, the states with ${\cal M}=0$ are not degenerate, while
the others with ${\cal M} \ne 0$ are doubly degenerate. If $n$ is odd,
the states with ${\cal M}=0$ do not exist and, therefore, all the eigenstates
of $\hat H^{(0)}$ are doubly degenerate.
If $U_a<0$, the ground-state energy
$
  E_g(\psi_0,\psi_\Lambda)
  =
  E^{(0)}_{n/2,\pm n/2}
$,
and the chemical potential $\mu$ varies in the interval
\begin{displaymath}
n-1
<
\frac{\mu}{U_s-|U_a|}
<
n
\;,\quad
n=1,2,\dots
\end{displaymath}
In the case of $U_a>0$,
$
  E_g(\psi_0,\psi_\Lambda)
  =
  E^{(0)}_{n/2,\pm 1/2}
$
and
\begin{displaymath}
U_s(n-1)<\mu<U_s n - U_a
\;,
\end{displaymath}
if $n$ is odd.
If $U_a>0$ and $n$ is even,
$
  E_g(\psi_0,\psi_\Lambda)
  =
  E^{(0)}_{n/2,0}
$
and
\begin{displaymath}
U_s(n-1)-U_a<\mu<U_s n
\;.
\end{displaymath}
The minimum of $E_g$ in all the cases is reached at $\psi_0=\psi_\Lambda=0$
and the system has a vanishing compressibility; i.e., it is in
the Mott-insulator phase.
Since
$\langle n/2,{\cal M}| \hat T_3 |n/2,{\cal M}\rangle=0$,
if ${\cal M}$ is the isospin projection on the direction $1$ or $2$,
the mean occupation numbers of the componets
$
  \langle n/2,{\cal M}| \hat n_0 |n/2,{\cal M}\rangle
  =
  \langle n/2,{\cal M}| \hat n_\Lambda |n/2,{\cal M}\rangle
  =
  n/2
$.

In order to investigate the transition from the Mott phase into the superfluid phase,
one has to calculate the ground-state energy of the complete Hamiltonian (\ref{HBH1}).
This can be done exactly by numerical calculations or approximately treating
the tunneling $\hat V$ as a perturbation. Let us consider the case $U_a<0$ first.
Nonvanishing matrix elements of the operator $\hat V$ are given by
\begin{eqnarray}
\left\langle \frac{n-1}{2},{\cal M} \mp \frac{1}{2} \right|
\hat V
\left| \frac{n}{2},{\cal M} \right\rangle
&=&
-
d J
\sqrt{n \pm 2{\cal M}}
\\
&\times&
\left(
    \psi_\Lambda
    \pm
    \psi_0
\right)
\exp
\left(
    \mp i \pi / 4
\right)
,
\nonumber
\end{eqnarray}
which implies that the ground-state energy $E_g$
is independent of the sign of $\psi_\sigma$,
$
E_g(\psi_\Lambda,\psi_0)=E_g(|\psi_\Lambda|,|\psi_0|)
$,
and it is a symmetric function of
$\psi_0$ and $\psi_\Lambda$: $E_g(\psi_0,\psi_\Lambda)=E_g(\psi_\Lambda,\psi_0)$.
The minima of $E_g$
are located on the lines $\psi_\Lambda=\pm \psi_0$~\cite{KGPRA04}.
Therefore, for the calculation
of the phase diagram of the system one can set $\psi_\Lambda=\pm \psi_0 \equiv \psi$.
Employing the perturbation theory we found that $E_g$ has the structure
\begin{equation}
\label{pert}
E_g(\psi)
=
a_0
+
a_2 \psi^2
+
a_4 \psi^4
+
\dots
,
\end{equation}
where $a_2$ can be either positive or negative,
but $a_4$ is always positive, which means that the SMQPT is
second order. The boundary between the Mott and superfluid phases is
determined from the condition $a_2=0$ and given by
\begin{equation}
\label{pbn}
\tilde J
=
\left(
    1-n+\tilde\mu
\right)
\left(
    n-\tilde\mu
\right)
/
\left(
    1+\tilde\mu
\right)
\;,
\end{equation}
where $\tilde J = 2 d J/(U_s-|U_a|)$ and $\tilde\mu=\mu/(U_s-|U_a|)$,
similar to the case of spinless bosons~\cite{Sachdev,Oosten}.
This equation agrees perfectly with the results obtained by numerical
diagonalization.

\begin{figure}[t]
\centering

\psfrag{e}[br]{$E_g/U_s$}
\psfrag{p}[b]{$\psi$}

  \includegraphics[width=5cm]{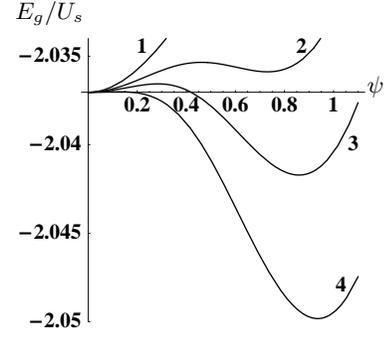}

\caption{Ground-state energy of the Hamiltonian (\ref{HBH1}) for $^{23}$Na
         ($U_a/U_s \approx 0.037$~\cite{Ho}).
         $\mu/U_s=1.5$, $2 d J/U_s=0.125(1),0.148(2),0.157(3),0.167(4)$.
        }
\label{egp}
\end{figure}

\begin{figure}[t]
\centering

\psfrag{m}[l]{$\mu/U_s$}
\psfrag{P}[b]{$U_a/U_s$}

  \includegraphics[width=5cm]{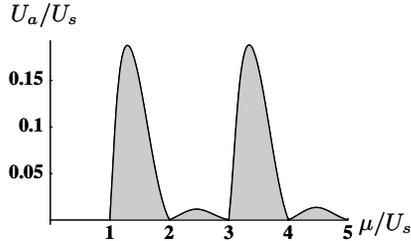}

\caption{In the shaded regions of this diagram, $a_4$ is negative and
         $E_g(\psi)$ has two minima at certain values of $J$.
	 In the remaining part $a_4$ is positive and $E_g(\psi)$ has only one minimum.
        }
\label{pmu}
\end{figure}

\begin{figure}[t]
\centering

\psfrag{J}[l]{$2dJ/U_s$}
\psfrag{m}[b]{$\mu/U_s$}

\psfrag{MM}[lb]{ MS}
\psfrag{MS}[lb]{ MM}

  \includegraphics[width=6cm]{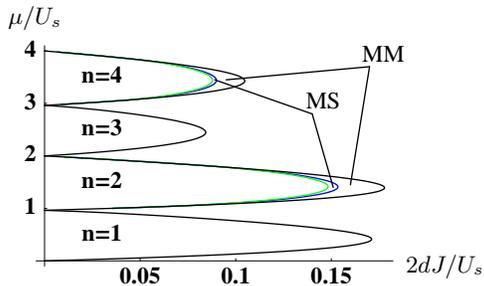}

\caption{Phase diagram for $^{23}$Na ($U_a/U_s \approx 0.037$~\cite{Ho}).
         The regions of metastable superfluid phase coexisting with the stable
	 Mott phase and that of metastable Mott phase coexisting with
	 the stable superfluid phase are denoted by MS and MM, respectively.
        }
\label{pdp}
\end{figure}

In the case of positive $U_a$, the situation is quite different.
Since the matrix elements of the operator $\hat V$ are given by
\begin{eqnarray}
\left\langle \frac{n-1}{2},{\cal M} \mp \frac{1}{2} \right|
\hat V
\left| \frac{n}{2},{\cal M} \right\rangle
&=&
-
d J
\sqrt{n \pm 2{\cal M}}
\\
&\times&
\left(
    \psi_0
    e^{i \pi / 4}
    \mp
    \psi_\Lambda
    e^{-i \pi / 4}
\right)
\;,
\nonumber
\end{eqnarray}
the ground-state energy $E_g$ depends only on $\psi^2=\psi_0^2+\psi_\Lambda^2$.
Typical dependences $E_g(\psi)$ obtained by numerical diagonalization are shown
in Fig.~\ref{egp}. In contrast to the case of negative $U_a$, there can be one or two minima.
This leads to the fact that the QPT can be not only first order but also second
order and the superfluid and Mott phases can coexist in certain parts of the
phase diagram. Analytical calculations within the framework of the perturbation
theory show that $E_g$ has the form (\ref{pert}), but now not only $a_2$ but also
$a_4$ can be either positive or negative. Therefore, in order to be able to work out
the complete phase diagram analytically, one has to calculate $a_6$ in Eq.(\ref{pert}).
This is a tedious task and we have not done that. However, solving the equation
$a_4(\mu/U_s,U_a/U_s)=0$ one can determine the regions
in the plane $(\mu/U_s,U_a/U_s)$, where $E_g$ can have one and two minima; i.e., one can
find the regions where the SMQPT is second and first order, respectively. These regions,
which are shown in Fig.~\ref{pmu},
can be also found by numerical diagonalization and the result appears to be the same.
Our analytical as well as numerical calculations show that, for $n=1$, $E_g(\psi)$ has
only one minimum and the SMQPT is always second order. If $n \ge 2$, $E_g(\psi)$ can
have either one or two minima, depending on the parameters, and the SMQPT can be either
second or first order. If, at a fixed $n$, $U_a/U_s$ is larger than some critical value,
which is about $0.188$ for even $n$ and grows from $0.012$ ($n=3$) to $0.015$ ($n\to\infty$)
for odd $n$, the QPT is second order; otherwise, it is first order.
In the case of $^{23}$Na shown in Fig.~\ref{pdp}, an interesting regime is achieved,
when the QPT for odd $n$ is second order, but for even $n$ it is first order.

The phase diagram for $^{23}$Na presented in Fig.~\ref{pdp}
has been obtained by numerical diagonalization.
It consists of a series of (internal) lobes corresponding to the stable Mott phase
and external regions corresponding to the stable superfluid phase. However,
in the case of even $n$, the two regions are separated from one another by intermediate
ones, where the stable and metastable superfluid and Mott phases coexist.
The boundary separating the region of the stable superfluid phase from other
ones can be determined from the condition $a_2=0$. If $n$ is odd, it is given by
\begin{eqnarray}
&&
J'
=
4
\left[
    \frac{n-1}{\mu'-n+1+2U_a'}
    +
    \frac{n+3}{n+U_a'-\mu'}
\right.
\nonumber\\
&&
\left.
    +
    2
    (n+1)
    \left(
        \frac{1}{n-U_a'-\mu'}
	+
        \frac{1}{\mu'-n+1}
    \right)
\right]^{-1}
\;,
\end{eqnarray}
where $J'=2 d J/U_s$, $U_a'=U_a/U_s$, and $\mu'=\mu/U_s$.
For even $n$, the boundary is given by the equation
\begin{equation}
J'
=
\frac
{
 \left(
     \mu'-n+1+U_a'
 \right)
 \left(
     n-\mu'
 \right)
}
{
 \mu'-n+1+U_a'
 +
 \left(
     1+U_a'
 \right)
 n/2
}
\;.
\end{equation}
These analytical expressions describe perfectly the corresponding numerical results.

In conclusion, we have shown that in the laser configuration we consider,
the SMQPT of spin-1 bosons
with the ferromagnetic interactions ($U_a<0$) is always second order and the boundary
between the superfluid and Mott-insulator phases is given by Eq.(\ref{pbn}).
In the case of antiferromagnetic interactions, the SMQPT is first order
in the shaded region of the plane $(\mu,U_a)$, shown in Fig.~\ref{pmu},
and it is second order in the rest part of the plane.

\begin{acknowledgments}
This work has been supported by the SFB/TR 12
``Symmetries and universalities in mesoscopic physics".
\end{acknowledgments}



\begin{thebibliography}{99}
\bibitem{FO}
First-order SMQPT's in a system of spin-1 bosons but in a different laser configuration
were also found recently in Ref.~\cite{KTK}.
\end{thebibliography}
\end{document}